\documentclass[12pt,preprint]{aastex}
\usepackage{psfig}
\received{2003 February 7}
\begin{document} 
\def\etal{{\it et al.\/}} 
\def\cf{{\it cf.\/}} 
\def\ie{{\it i.e.\/}} 
\def\eg{{\it e.g.\/}}

\title{On the generation of UHECRs in GRBs: a reappraisal} 
\author{{\bf Mario Vietri$^1$, Daniel De Marco $^2$ and Dafne Guetta$^3$}} 
\affil{$^1$ Scuola Normale Superiore, Pisa\\ 
$^2$ Universit\'a di Genova \\
$^3$ Osservatorio Astrofisico di Arcetri, Firenze 
} 
{} 
\begin{abstract} 
We re-examine critically the arguments raised against the theory that 
Ultra High Energy Cosmic Rays observed at Earth are produced in  
Gamma Ray Bursts. These include the limitations to the highest energy 
attainable by protons around the bursts' shocks,  
the spectral slope at the highest energies,  
the total energy released in non--thermal particles,  
the occurrence of doublets and triplets in the 
data reported by AGASA. We show that, to within the uncertainties in
our current knowledge of GRBs,  none of these objections is really 
fatal to the scenario. In particular, we show that the
total energy budget of GRBs easily accounts for the
energy injection rate necessary to account for UHECRs
as observed at Earth. We also compute the expected particle spectrum
at Earth, showing that it fits the HiRes and AGASA data to 
within statistical uncertainties. We consider the
existence of multiplets in AGASA' data. To this end, 
we present a Langevin--like treatment for the motion of
a charged particle in the IGM magnetic field, which allows us to 
estimate both the average and the rms timedelay for particles of
given energy; we discuss when particles of identical
energies reach the Earth in bunches, or spread over the rms
timedelay, showing that multiplets pose no problem for an
explosive model for the sources of UHECRs. 
We compare our model with a scenario
where the particles are accelerated at internal shocks,
underlining differences and advantages of particle acceleration at  
external shocks.
\end{abstract}

 
\section{Introduction} 
 
We suggested some time ago (Vietri 1995) that Gamma Ray Bursts provided 
the ideal accelerating sites for UHECRs, supporting this view with  
a computation of the highest energy attainable by protons accelerated  
around the fireball's external shock (which was found to exceed  
$10^{20} eV$), and of the average energy deposited by bursts per 
unit time and volume in gamma ray photons (which came close to that  
inferred for UHECRs, if these have a cosmological origin); assuming 
comparable efficiencies in the production of gamma ray photons  
and non--thermal particles, we argued that this provided an excellent 
explanation, without fitting any parameter, of the UHECRs' flux 
levels at Earth. Since then, our knowledge of bursts has increased 
dramatically, and the fireball theory (Rees and Meszaros 1992)  
has received a spectacular confirmation with the accurate description 
of the properties of afterglows (see Piran 1999, 2000 for reviews). 
Independently of this,  
the recently reported data from HiRes (Abu--Zayyad {\it et al}., 2002) 
appear once again to support the view that the origin of the Ultra High  
Energy Cosmic Rays (UHECRs), at least as observed at Earth, is to be  
found among astrophysical sources. These two facts together make it 
worthwhile to re-examine 
the objections that were raised against the above--mentioned scenario, 
in the wake of its first appearance. Also, an alternative model for the  
production of UHECRs in GRBs has been proposed (Waxman 1995), and thus it 
seems pertinent to discuss some differences between these two models,  
which will be subject to different observational tests. 
 
The plan of the paper is as follows: in Section 2, we address the  
criticisms levied against the model in the past years which concern 
the acceleration mechanism. In Section 3, we deal with global
energetics. We discuss in Section 4 two caveats on the
previous discussion on energetics. Doublets and triplet in the data 
reported by AGASA,  are discussed in Section 5; we compare our work to 
that of other  authors in Section 6, then we summarize our conclusions. 
 
\section{Acceleration} 
 
The details of the model for the acceleration of non--thermal particles 
remain essentially those outlined in Vietri (1995), and in Dermer (2002a):  
particles are  
accelerated around the fireball's external shock, which is highly  
relativistic for most of the relevant time. The major reason for this 
is that the basic fireball model employed (Meszaros, Laguna, 
and Rees, 1993, where a detailed model for the external shock evolution 
is presented, together with the physical characteristics of the plasma behind  
the shock) has proved itself able to explain the 
characteristics of the afterglow (Piran 1999, 2000), and it thus appears in need 
of no major revision. The model was used mostly in determining the largest 
energy which could bounce off the ejecta shell, without crossing it 
unscathed (and also, in showing that no radiative losses 
where incurred into, by the highest energy particles). We derived for the 
highest energy 
\begin{equation} 
\label{Emax}
E_{max} = 10^{20} eV E_{51}^{1/3} \Gamma_2^{1/3} n_1^{-5/6}  
\epsilon_B^{1/2} \frac{1}{1-\cos\theta} 
\end{equation} 
which differs from the original (Eq. 34 in Vietri 1995) because a more 
modern initial Lorentz factor $\Gamma = 100$ has been used,  
the true energy release $E$ and beaming angle $\theta$ have been introduced.
The above equation defines the largest energy which can be turned back 
in a typical wind which produces GRBs, as seen in the upstream frame 
(the laboratory); the very important question of why we should see 
particles leaving the shock region through the upstream frame rather than 
the more customary downstream frame will be addressed shortly; let us
just remark that particles leaving through the downstream region will
be seen in the laboratory frame as having an energy roughly $\Gamma^2$
smaller than the above.

The average length for deflecting back a non--thermal particle has been 
taken as $\approx 40$ times its Larmor radius, but the computation holds 
only for $\Gamma \gg 1$. It was shown in Vietri (1995) that losses due to all
radiative processes are negligible. Furthermore, no particle loss 
may occur through the sides of the ejecta (which would be conceivable,
because the emission is beamed along a cone of semiopening angle $\theta$);
in fact, in the comoving frame, the ejecta have a (radial) width given
by $r/4\Gamma$, where $r$ is the shock's distance from the explosion site,
while the ejecta width in the transverse direction is $\approx r\theta$,
and, since $1/\Gamma < \theta$ (Panaitescu and Kumar 2002) it will be
easier for the particles to leave through the bottom end than through
the sides. Recently, Panaitescu and Kumar (2002) have obtained values for 
the circumburst's 
density $n$, for the departures of the magnetic field from its  
equipartition value $\epsilon_B$, for the total energy $E$ release and 
beam opening angle $\theta$, 
for a variety of bursts's afterglows for which sufficient time and frequency   
coverage were available; using the data in their Tables 2 and 3, we compile  
our Table 1, illustrating the largest energies achievable within this scenario, 
which are conveniently large with respect to observational data. 

\clearpage
 
\begin{table} 
\begin{center} 
 {\bf TABLE 1.}  
   Largest energy achievable by protons accelerated at external shocks,
   and particle spectral index.\\ [4ex] 
\begin{tabular}{cc} 
 \hline \hline 
\rule[-2.5mm]{0mm}{8mm}{} 
  GRB  & $E_{max}/10^{20}\; eV$  \\ 
 \hline 
\rule[-2.5mm]{0mm}{8mm} 
  970508 & $7.2$  \\ 
\rule[-2.5mm]{0mm}{8mm} 
  980519 &  $38.0$   \\ 
\rule[-2.5mm]{0mm}{8mm} 
  990123 &   $5740.$   \\ 
\rule[-2.5mm]{0mm}{8mm} 
  990510 & $161.$   \\ 
\rule[-2.5mm]{0mm}{8mm} 
  991208 & $1.80$   \\ 
\rule[-2.5mm]{0mm}{8mm} 
  991216 & $18.$    \\ 
\rule[-2.5mm]{0mm}{8mm} 
  000301c &  $0.5$  \\ 
\rule[-2.5mm]{0mm}{8mm} 
  000418 &   $0.2$   \\ 
\rule[-2.5mm]{0mm}{8mm} 
  000926 &  $1.4$  \\ 
\rule[-2.5mm]{0mm}{8mm} 
  010222 &  $ 1.3$ \\ 
 \hline \hline 
\end{tabular} 
\end{center} 
\end{table} 
 
\clearpage 
 
A critique against this argument was levied by Gallant and 
Achterberg (1999), who pointed out that the limiting factor for the 
highest energy achievable is due to the ability of the insterstellar medium 
to deflect high--energy particles toward the shock, and not the  
{\it vice versa}, which is what was computed above. Their analysis showed that  
this highest energy was 
\begin{equation} 
\label{gallant} 
E_{max} \approx 10^{15}\;eV \frac{B}{1 \mu G}\;. 
\end{equation} 
This argument however assumes a definite value for the magnetic field around 
the burst, which is very model--dependent: there is no compelling reason 
to assume for $B$ the value of the average ISM. In particular, two of the leading  
models for GRBs, the binary neutron star merger (Narayan, Paczynski and Piran 
1992), and the SupraNova (Vietri and Stella 1998)  
are clearly at odds with this value. In fact, these two models make  
very similar predictions, because in both cases the GRB goes off inside a 
pulsar wind bubble (PWB, K\"onigl and Granot 2002; Inoue, Guetta and Pacini 2001,  
Guetta and Granot 2002, but the same, bare proposal was made also in Vietri 1995, 
and in Gallant and Achterberg 1999), filled with an  
electron/positron gas, and its associated  
magnetic field. In either case, a pulsar--type dipolar magnetic field 
($B = 10^{12}-10^{13}\;G$) has a light cylinder at $r_{lc} \approx 10^7\; cm$, 
and decreases like $B \propto r^{-1}$ (at least in its toroidal component),  
from there on, as dictated by standard pulsar electrodynamics (Contopoulos,  
Kazanas and Fendt, 1999). This results in fields  
of order $B = 0.1-10\; G$ at the radii appropriate for afterglows  
($r = 10^{16}-10^{17}\; cm$). These estimates are in agreement with  
determinations of the magnetic field of the (much slower rotating!) Vela 
bubble  (Helfand, Gotthelf and Halpern 2001, but see Arons 2002 for a  
different estimate of the magnetic field). 

In order to adapt Eq. \ref{gallant} to our needs, we repeat Gallant and 
Achterberg's computations
for $B \propto r^{-1}$. We consider a particle leaving the shock at radius $R_0$ 
with Lorentz factor $\gamma$, and moving initially perpendicular to the shock,
in the observer's frame; the shock has Lorentz factor $\Gamma_s = \Gamma_s(R) 
\gg 1$, so that most particles will be emitted perpendicular to the shock 
(Bednarz and Ostrowski 1998). 
We also have allowed for the shock deceleration. Because of a component of the
magnetic field $B_\perp (R) = B_1 R_1/R$ perpendicular to the shock normal,
and thus to the particle initial direction of motion, the particle speed
perpendicular to the shock $v_z$ changes according to 
\begin{equation}
\frac{v_z}{c} \approx 1 - \frac{e }{m \gamma c^2}\int_{R_0}^R B_\perp d\!R = 
1- \frac{e B_1 R_1}{m \gamma c^2}\ln(R/R_0)\;.
\end{equation}
The shock moves instead with speed $v_{shock}/c \approx 1- 1/2\Gamma_s^2$.
It is necessary that the shock overcomes the particle promptly; only in this
case, in fact, the returning particle will see a shock with the same 
Lorentz factor as when it left the post--shock region, and conventional
Fermi--Type acceleration will take place; if this were not the case, we 
would have a strongly evolutionary shock, and the ordinary predictions 
of steady--state theory would not follow (and, most likely, the shock would
not be an efficient particle accelerator). We must thus have $v_{shock} d\!R 
> v_z d\!R$, or, using the equations above, 
\begin{equation}
\ln(R/R_0) > \frac{m \gamma c^2}{2 \Gamma_s^2 e B_1 R_1}
\end{equation}
where $R$ is the radius at which the particle recrosses the shock.
Demanding again that the shock promptly overcomes the particle means that
$\ln(R/R_0) \approx \delta\!R/R_0 \ll 1$, and thus we find that only 
particles with 
\begin{equation}
E_p = m\gamma c^2 < 2 \Gamma_s^2 e B_1 R_1 \equiv E_{mb}
\end{equation}
are accelerated by the usual Fermi mechanism. Particles with $E>E_{mb}$
will find a weaker, slower shock if they come back, and will not be
significantly accelerated any more. Because of Eq. \ref{adiabatic}, we
also see from the above that the condition that particles never come
back (assuming that the PWB merges into the ISM field for $R \gtrsim
10^{18}\; cm$) is only marginally stronger than the above: we find
\begin{equation}
\label{released}
E_p \approx E_{mb} = 2 \Gamma_s^2 e B_1 R_1\;.
\end{equation}

Gallant and Achterberg (1999) used $R_1 = R_d = 10^{16}\; cm$, the approximate 
radius at which the afterglow begins, the typical ISM field $B = 1\;
\mu G$, and the initial shock Lorentz factor to obtain Eq. \ref{gallant}.
But for a PWB, following the discussion above we may take for the field 
outside the light cylinder radius (Eq. 31 of Goldreich and Julian 1969): 
\begin{equation}
B(R_1) R_1 = 10^{17} \; G \; cm 
\frac{B}{5\times 10^{12}\; G} \left( \frac{1 \; ms}{P}\right)^2
\sin\theta\frac{\Psi(\theta)}{I}\;.
\end{equation}
Here $P$ is the pulsar's rotation period, which we have scaled to $1\; msec$, which 
is about right for both neutron star mergers and SupraNovae; we took for the dipolar
field of the neutron star a standard value, $\theta$ is the beams' semiopening angle
and $\Psi(\theta)/I$ is a normalized fudge factor which Gooldreich and Julian (1969)
do not estimate. However, inspection of Fig. 3 of Contopoulos, Kazanas and Fendt (1999)
shows that $\Psi(\theta)/I \approx (\sin\theta)^2$; using $\theta\approx 0.1$ from
Frail {\it et al}. (2001) and Panaitescu and Kumar (2002), we find 
\begin{equation}
B(R_1) R_1 \approx 10^{14}\; G \; cm \;,
\end{equation}
independent of $R_1$. Strictly speaking, we have $100 \gtrsim \Gamma_s \geq 1$,
but for a Newtonian shock particle acceleration proceeds on a slower, diffusive
timescale (Meli and Quenby 2002; see also the discussion later in this 
section). From Fig. 2 of Kirk {\it et al}. (2000), we see that the particle 
spectral index departs from it asymptotic value for very large shock Lorentz
factors when $\Gamma_s \approx 2$. So we shall use $100 \gtrsim \Gamma_s \geq 1$,
but keeping this caveat in mind:
\begin{equation}
10^{17} \; eV \leq E_{mb} \leq 6\times 10^{20}\; eV\;,
\end{equation}
depending on the instantaneous value of $\Gamma$. Since we know that
(Piran 2000)
\begin{equation}
\label{adiabatic}
\Gamma_s^2 = \Gamma^2 \left( \frac{R_d}{R}\right)^3
\end{equation}
where $\Gamma_s \approx 100$ (Panaitescu and Kumar 2002) is the initial 
shock Lorentz factor, we see that $E_{mb}$ spans continuously the range 
above, as $R$, the instantaneous shock position, changes. 

We thus have the following picture. So long as $E_{mb} > E_{max}$, particles
cannot escape in the upstream direction: they are promptly overtaken by the
shock because they are effectively kicked backwards by
the environment magnetic field, and continue their acceleration cycle
until they reach 
an energy $E_{max}$. When this occurs, the particles cannot be effectively 
deflected by the ejecta's magnetic field: they will cross unhindered 
the hyperrelativistic shell, and be lost through the back of the shell. 
When this occurs, their energies in the laboratory (upstream) frame are 
a full factor $1/\Gamma^2$ lower than given by Eq. \ref{Emax}, which 
holds {\bf only }for particles leaving the shell in the forward direction.
However, when $E_{mb} < E_{max}$, particles which have been accelerated
up to $E_{mb}$ (in the laboratory=upstream frame) cannot be deflected
backwards toward the shock any longer, and will be able to escape in the
upstream direction without further acceleration. The critical moment
occurs when $E_{mb} = E_{max}$ (where now in Eq. \ref{Emax} the shell Lorentz
factor must be taken to vary as in Eq. \ref{adiabatic}): equating the two, 
we find that $\Gamma \approx 30$, and thus that $E_{max}$ at this moment is a 
factor $1.5$ lower than Eq. \ref{Emax} (and thus, within the accuracy of 
these estimates, it is identical to Eq. \ref{Emax}). We thus have
that the energy range for particles escaping in the upstream direction is
\begin{equation}
\label{range}
10^{17}\; eV \leq E \leq E_{max}\;.
\end{equation}
It should be pointed out that the above picture differs from Gallant and 
Achterberg's (1999).
{}

We now have to derive the spectrum of these particles. Here there are two
distinct possibilities. On the one hand, one may take Vietri's (1995) and
Gallant and Achterberg's (1999) view that the particles which end up as UHECRs
are the relativistic particles pre--existing the burst which must necessarily
exist in a PWB. Under this assumption, Gallant and Achterberg showed that
the spectrum will be of the form
\begin{equation}
d\!n(E) \propto E^{-2} d\!E\;.
\end{equation}

On the other hand, one may assume that the total number of energetic particles
is dwarfed by the (formerly thermal) particles injected at the shock, in
which case the spectrum of particles can be derived as follows.
Assume that a fraction $q$ of the total fireball energy $E_f$ is converted
on the timescale $t_{acc}$ in particles of energy given by Eq. \ref{released}.
There will then be a total amount of energy
\begin{equation}
\frac{q E_f}{t_{acc}} d\!t = \frac{q E_f}{2}\frac{d\!R}{\Gamma_s R}
\end{equation}
in particles of energy $\approx$ given by Eq. \ref{released}. Eliminating
$\Gamma_s$ and $R$, and using (Gallant and Achterberg 1999)
\begin{equation}
t_{acc} \approx t_u = \frac{E_p}{\Gamma_s e B c}
\end{equation}
where $t_u$ is the upstream residence time, we find a particle spectrum
\begin{equation}
d\!n \propto E^{-5/2} d\!E\;,
\end{equation}
Putting together all of the above, we shall take a spectrum of the form
\begin{equation}
\label{spectrum1}
d\!n \propto E^{-\gamma}\;\;;\;\; 2\leq\gamma\leq 2.5\;.
\end{equation}
We shall then let $\gamma$ be a free a parameter in the range above, and use
it to fit observations, in the next section.

It is worth remarking at this point that the above results depends upon our 
having assumed that the GRB goes off inside a PWB (K\"onigl and 
Granot 2002), a natural assumption when GRBs are due to netron star mergers
or to SupraNovae. While currently neutron star binary mergers are not favored as 
an explanation for long--duration GRBs, they appear attractive as a model
for short GRBs; while  SupraNovae are one of the leading contenders for the long 
bursts (see Dermer 2002b for a review). But the above discussion applies, at
least qualitatively as a lower limit, to a wide class of models, including
Usov's (1992) model of a fast pulsar with exceptionaly strong magnetic field, 
models based upon ADIOS flows (Blandford and Begelamn 1998,Blandford 2002), 
Spruit's models based on MHD winds (Drenkhahn and Spruit 2002, and references 
therein), or several distinct models all involving magnetized neutron stars
(Duncan and Thompson 1992, Dai and Lu 1998, Blackman and Yi 1998). 

One may wonder whether losses in the somewhat more photon- and magnetic field-
rich environment provided by a PWB make the estimate of Eq. 1 unrealistically
optimistic. This was studied by Inoue, Guetta and Pacini (2002), who proved
this to be the case only under exceptional circumstances. It was in fact their
aim to explain the high--energy ($> 1\; GeV$) emission occasionally seen by
EGRET in GRBs, and this could be achieved only with rather extreme assumptions
on the environment parameters. This is not at all in contradiction with 
the above discussion: $GeV$ emission in GRBs is a rare phenomenon, so it makes
sense that models to explain it require extreme values for the parameters in
play.

A different critique was levied (Ostrowski and Bednarz 2002) concerning the 
spectral slope for particles accelerated at highly relativistic shocks. These 
authors pointed out that the claims of universality for the spectral index  
$s \approx 4.2$ (Kirk and Schneider 1987, Heavens and Drury 1988, Bednarz 
and Ostrowski 1998, Kirk {\it et al}., 2000) possibly overstate their case, 
because they neglect the fact that, as the regular magnetic field in the  
downstream region becomes stronger than the field associated with the turbulent 
component, the spectral slope must increase, the spectrum softens, and the 
claim to universality evaporates. This criticism, however, can be 
countered in two (orthogonal) ways. On the one hand, it is possible that 
it applies to a somewhat idealized situation, whereby post--shock material is 
incapable of generating a small--scale magnetic field, and thus the  
field compressed at the shock dominates the turbulent component. This is the 
view of Medvedev and Loeb (1999) or Thompson and Madau (2000), who proposed 
ways in which magnetic field of non--negligible amplitude could be generated 
around the shock, and would be intrinsically turbulent. On the other hand,  
one may take the view of K\"onigl and Granot (2002), according to whom the  
very large values of $\epsilon_B$ derived for some bursts (in particular, 
GRB 970508 has $\epsilon_B = 0.1$ as determined by several independent studies: 
Wijers and Galama 1999, Granot, Piran 
and Sari 1999, Chevalier and Li 2000, Panaitescu and Kumar 2002) cannot 
possibly be accounted for in this way, but are instead easily accounted 
for by the compression of a magnetic field surrounding the burst explosion 
site, provided this occurs in a PWB, which again 
occurs naturally if bursts are due to binary neutron star mergers, or  
SupraNovae. The dominant component of this magnetic field is of course 
toroidal (Goldreich and Julian 1969), and its sweep up and compression  
would give a post--shock magnetic field mostly oriented along the shock surface, 
which would increase, not decrease, the shock effectiveness in the acceleration 
of particles.  
 
Another criticism was that the acceleration time scale was shorter than 
hypothesized in Vietri (1995), because the average energy gain $<G>$ does  
not scale as $\Gamma^2$, where $\Gamma$ ($\approx 100$, Panaitescu and Kumar  
2002) is the shock Lorentz factor. It was pointed out that this 
energy gain applies to the first shock crossing, but the following 
crossings would provide a more mundane $<G> \approx 2$ (Gallant and Achterberg  
1999). This criticism is surely correct, but does not spoil the gist of the 
original argument, that the acceleration time scale is much shorter than for  
Newtonian shocks. Indeed, the fact 
that the energy gain per shock crossing is no longer infinitesimal, as it  
is for Newtonian shocks (where $G-1 = v_{shock}/c \ll 1$), implies that  
the required number of shock crossings for acceleration to the EeV range 
becomes finite: from $m \Gamma^2 c^2$ to the EeV range only $\approx  20$ 
shock crossings are required,  
in agreement with the cosiderable reduction of the acceleration time scale  
seen in numerical simulations (Meli and Quenby 2002). This very fact has led  
some of the severest critics of the scenario to propose themselves that 
UHECRs are accelerated at highly relativistic shocks (Kirk, Achterberg, 
Guthmann and Gallant 2000).  
 
There is a deep reason for this: we showed elsewhere (Vietri 2002) that, 
for particle acceleration around shocks of any speed, the following relation 
holds: 
\begin{equation} 
P <G^{3-s}>  = 1\;. 
\end{equation} 
Here $P$ is the probability of returning to the shock, $G$ the energy  
amplification for one cycle across the shock, $s$ the distribution function 
spectral index, and the average is to be taken after the raising to the  
$3-s$ power; the above generalizes to relativistic shocks Bell's (1978) 
relation for the spectral index. The above relationship holds irrespective 
of the assumed scattering law. In the case of Newtonian shocks, both $P$ 
and $G$ differ from unity by quantities ${\cal O}(v/c) \ll 1$, and the two 
combine to give $s = 4$. In the hyperrelativistic case, both $P$ and $G$ 
differ significantly from unity, so that, since $P \approx 1/2$ (the  
post--shock fluid runs away from the shock with speed $\approx c/3$, 
so that it is easy for particles to be advected away), also $G-1$ 
must grow away from its infinitesimal Newtonian values, and tends to 
$G \approx 2$. This, of course, shortens the acceleration 
time scale, and is a robust prediction of the theory of acceleration 
around hyperrelativistic shocks, independent of the scattering law.  

\section{Energetics}

Another criticsm which has been levied against this scenario is that,  
contrary to our original claims, there is a mismatch between the observations 
of the energy injection rate in $\gamma$--ray photons due to GRBs, and in  
UHECRs. This claim has been made independently by Stecker (2000) and  
Berezinsky {\it et al}. (2002), who pointed out that the realization that 
the redshift distribution of GRBs should follow the star formation history 
in the Universe must lower the local energy injection rate. In particular,
Stecker found the evolving GRB population to produce a $\gamma$ ray photon
flux which is an order of magnitude lower than that of just the fraction of
UHECRs beyond $10^{20}\; eV$, let alone the rest of the spectrum.. 
Waxman (2002)  made some quite reasonable criticisms of the paper by
Berezinsky {\it et al}. (2002). We consider here Stecker's work. We shall
first show that Stecker's claims are not supported by currently available
data, then derive the spectrum of UHECRs if these are emitted by GRBs, and
then comment on the (rather large) uncertainties still surrounding the 
overall energetics of GRBs.  

Stecker's (2000) estimates are based upon old data: 
he used the AGASA flux at $E > 10^{20}\; eV$ (Takeda {\it et al}., 1998, 
Hayashida {\it et al}., 1999) which he took to be an unbroken
power--law with index $2.78$ beyond $10^{20}\; eV$, while Abu--Zayyad 
{\it et al}. (2002) explicitly state that their data beyond $10^{19.8}\; eV$ 
depart from such a law, and state that the number of events they observe is a 
full factor of $4/19.1 \approx 0.21$ below the extrapolation of 
such law. Thus, while he took the local energy release rate in UHECRs to be
\begin{equation}
\dot{\epsilon}_{(AGASA)} = 2\times 10^{44} \; erg\; Mpc^{-3}\; yr^{-1}\;, 
\end{equation}
HiRes data suggest a value a factor of $0.21$ lower:
\begin{equation}
\label{rateuhecrs}
\dot{\epsilon}_{(HiRes)} = 4\times 10^{43}\; erg \; Mpc^{-3}\; yr^{-1}\;.
\end{equation}
which compares already much better with the value he took for the energy 
release rate in $\gamma$--ray photons, $2.2\times 10^{43}\; erg\; Mpc^{-3
}\; yr^{-1}$. But even this value is superseded, since it is based on Schmidt's
(1999) estimate of the GRB rate. In his
new work, Schmidt (2001) gives the burst's rate (for isotropic bursts) as
\begin{equation}
\dot{n}_{GRB} = 0.5 \; Gpc^{-3}\; yr^{-1}\;.
\end{equation}
As correctly pointed out by Stecker (private communication), the above
value includes short bursts, about which, strictly speaking, we know very little. In particular, we ignore what their overall energy budget is, and, most important to this paper, whether they have relativistic afterglows at all.
So to play it safe, we reduce the above rate to exclude short bursts: 
According to Fishman and Meegan (1995) short bursts account for $\approx
30\%$ of all bursts, and we consequently take the local rate of long GRBs as
\begin{equation}
\dot{n}_{lGRB} = 0.35 \; Gpc^{-3}\; yr^{-1}
\end{equation} 
Schmidt does not give the average (isotropic) burst energy release, but we can
get this from Frail {\it et al.} (2001), who, among many other things, give the
isotropic burst energy release for all bursts with {\bf known redshifts}. The
reason why we prefer this over using Schmidt's work is that their total
energy releases (taken from Bloom, Frail and Sari 2001) cover the spectral
range $0.2-2000 keV$, while Schmidt restricts himself to the range
$50-300 keV$: there seems no obvious reason why we should adopt Schmidt's
values, when more suitable ones are available. From Frail et al.'s Table I, we find 
\begin{equation}
\label{photons}
E_{av} = 3.3\times 10^{53}\; erg\;.
\end{equation}
We thus find the energy release rate in $\gamma$--ray photons by GRBs as
\begin{equation}
\label{rategrbs}
\dot{\epsilon}_{GRB} = 1.1\times 10^{44}\; erg \; Mpc^{-3}\; yr^{-1}\;.
\end{equation}
Strictly similar results are obtained in an analogous way by using data
from Panaitescu and Kumar (2000) and from Piran {\it et al}. (2001). 
Please notice that, while not all data from these three sets of authors
are different (many bursts are in common) they fitted different quantities
with different methods, so the results are, for the most part, independent.
This is larger than $\dot{\epsilon}_{(HiRes)}$ by a hefty factor
of $\dot{\epsilon}_{GRB}/\dot{\epsilon}_{(HiRes)} = 3$, and allows a 
full explanation of UHECRs by GRBs, even in the case in which the efficiency
for particle acceleration is smaller than for photon production. 
 
We thus consider Stecker's (2000) objections superseded by new data. However,
we are still left with the question of how much energy is produced by GRBs in the
form of UHECRs in the whole energy range, Eq. \ref{range}, and not just
for $E > 10^{20}\; eV$, and whether the observed spectrum of UHECRs at Earth
can be fitted by a cosmological distribution of GRBs. In order to answer both of
these questions, we repeat here a computation of the spectra of UHECRs as 
observed at Earth,for the evolving population discussed above. This does {\bf not } 
mimick Waxman's (1995b) computations for several reasons. First, he uses a 
constant comoving density for GRBs, while, following Schmidt (2001), we use 
the second star formation history (SFR2) proposed by Porciani and Madau (2001); 
second, he uses a continuous energy loss approximation, while we take into
account the discreteness of the photopion production events (but not of 
pair production); and third, what we present here are {\it simulations}
of observations (De Marco, Blasi, Olinto 2003) in
order to stress that the fits to observations contain some amount of 
noise due to the finite number of events now available. For a full 
discussion of the method, simulations, and approximations, we refer the
reader to De Marco, Blasi, Olinto (2003). Following Eq. \ref{spectrum1},\ref{range}, 
the particle spectrum is taken as
\begin{equation}
\label{spectrum}
n(E) d\!E = E^{-\gamma}\; d\!E \;\; ; \;\; E_{min}\leq E \leq E_{max}
\end{equation}
where $\gamma$ is a free parameter that is to be varied to allow fitting
of Earth--based observations; $E_{min}$ is from Eq. \ref{range}, while
$E_{max}$ is irrelevant (provided it exceeds $10^{20}\; eV$, which it does,
Eqs. \ref{Emax},\ref{range}) because most energy is in the lowest energy
particles (for $\gamma > 2$). The spectrum we obtain is shown in Figs. 1
and 2 is obtained for the two values of $\gamma = 2.2$ and $2.5$. 
A smaller value of $\gamma$ yields only marginally different results.
The steeper spectrum, like $\gamma = 2.5$ fits data below $10^{19}\; eV$ 
much better, but, because of the obvious possibility of a Galactic 
contamination, we do not regard this accurate fit as either desirable or 
likely. We shall thus take the value
\begin{equation}
\label{slope}
\gamma = 2.2
\end{equation}
as our canonical result,
For both values of $\gamma$ we under-reproduce
the AGASA counts beyond $10^{20} \; eV$, but not the HiRes data.  

The local energy release rate in UHECRs, for the whole range of Eq. 
\ref{range} is 
\begin{equation}
\dot{\epsilon}_{UH} = 1.5\times 10^{45} \left(\frac{E_{min}}{10^{17}\; eV} 
\right)^{-0.2}\; \; erg\; Mpc^{-3}\; yr^{-1}
\end{equation}
which exceeds Eq. \ref{rategrbs} by a factor of $13$. This overestimate
is reduced when we consider the suggestion (De Marco, Blasi and Olinto
2003) that the energies of AGASA events are overestimated by a full $15\%$, 
which brings their data in much better agreement with HiRes'. In this
case, the total required energy release becomes
\begin{equation}
\label{uhecr}
\dot{\epsilon}_{UH} = 1\times 10^{45} \left(\frac{E_{min}}{10^{17}\; eV} 
\right)^{-0.2}\; \; erg\; Mpc^{-3}\; yr^{-1}
\end{equation}
which now exceeds Eq. \ref{rategrbs} by a factor of $9$ only. 

The estimate above is much lower than the claims by Scully and Stecker (2002), who
hypothesized an excess by a factor of $10^{2-3}$, over the much smaller
range $E> 10^{19}\; eV$, while the above equation applies to $E> 10^{17}
\; eV$. The origin for the disagreement between our paper and Scully
and Stecker's is twofold. First, we fit a different, flatter injection
spectrum ($2.2$ rather than $2.75$), but we still obtain excellent fits.
Second, we use a $\dot{\epsilon}_{GRB}$ (Eq. \ref{rategrbs}) which differs 
from theirs. However, they do not compute it anew, but they refer to Stecker 
(2000). We have already shown that this estimate is based upon old data, and 
will thus not discuss their results any further. 

The slight overabundance of energy required to explain UHECRs as observed at Earth
with respect to the energy injection rate in photons by GRBs, which
we estimated above as included in the range $10-13$, is not an embarrassment 
for the present model. Numerical simulations of radiative efficiencies
at internal shocks, the currently favoured model for the prompt phase of
GRB emission (Spada, Panaitescu and Meszaros 2000) show that these are
$\approx 0.005$, so that even the overproduction of UHECRs by a factor of
$9$ amounts to a total efficiency for UHECRs of a few percent, surely
in line with conventional thinking (Draine and McKee 1993)
about emission efficiencies for non--thermal particles around
non--relativistic shocks. For relativistic shocks, given the
shortening of the acceleration time--scale (Meli and Quenby 2002),
acceleration should be, if anything, more efficient. 

For this reason, and for the reasons to be discussed in the next paragraph,
we are not worried by this slight discrepancy.

\section{Two caveats}

There are two important caveats which need to be made when discussing bursts'
energetics.

Several authors have stressed that GRBs could have a redshift distribution
strongly skewed toward large redshifts, but while very likely, this
idea has not been confirmed yet. 
In order to stress this point, we have taken the set of all  
GRB redshifts known to date, to see whether the low redshift distribution 
agrees better with a  Porciani and Madau--like evolution law 
(Porciani and Madau 2001), or with a non--evolving 
one. We restrict our attention to small redshifts ($z < 1$) for two reasons: 
first, the Porciani and Madau evolution law is very distinctive at such  
low redshifts,  
and is not disputed by different authors; second, because it seems reasonable 
to assume that this sample is unbiased by the selection effects  
which obviously mar the higher end of the distribution.  
 
In Fig. 3 we show the observed cumulative distribution in redshift  
of known GRBs; data are taken from the compilation publicly 
available at  {\tt http://www.mpe.mpg.de/\~jcg/grbgen.html}, 
which also gives the relevant references. We compare it with  the
theoretical expectation that the GRBs explosion rate follows the star 
formation rate. 
We consider the three different star formation histories SF1, SF2 and SF3  
given by Porciani and Madau (2001). We use in the figure  
$\Omega_M=0.3$ and $\Omega_{\Lambda}= 0.7$ and a Hubble constant  
$H_0=65h_{65}$km s$^{-1}$ Mpc$^{-1}$. For completeness we compare the  
observed distribution also with a distribution $\propto (1+z)^3$. A KS 
test shows that the observed distribution without GRB 980425 differs  
from the theoretical ones SF1, SF2, SF3 and $\propto (1+z)^3$ with 
a significance of 24\%, 25\%, 28\% and 32\% respectively, 
while inclusion of GRB 980425 changes these values to 10\%, 11\%, 13\% and 
16\%.
 
Fig. 3 clearly shows that we have right now no compelling reason to believe 
that GRBs are characterized by very strong evolution with redshift, a 
uniform distribution providing an equally good, if not better, fit to current 
data. But furthermore, even if larger samples like the one to be collected  
by SWIFT should in the future provide a confutation of this (which we 
very much believe), still the point 
would remain that a perfectly normal subsample of a dozen or so objects 
may depart greatly from the average distribution, and appear like a  
uniform one. In this sense, then, the argument made by Stecker (2000) 
and Berezinsky {\it et al}. (2002) appears as an oversimplification of reality: 
in discussing sources of UHECRs, one cannot forget that the total number 
of sources one is dealing with is statistically very small, and likely to 
stay that way, and thus a proper discussion of the model on an energy basis  
should include the effects of fluctuations. 

Second, there is an important uncertainty in the 
global energetics of bursts, as we now discuss. Comparing the late--time
evolution of bursts (which fixes the total energy in the kinetic form)
with the energy released in $\gamma$--ray photons, it is possible 
to deduce the radiative efficiency for several individual bursts (Panaitescu
and Kumar 2000). The radiative efficiency thusly
derived is very high, always eceeding $66\%$ (except for
one case) and occasionally even approaching an incredible $90\%$ (see Table 
III of Panaitescu and Kumar). We remark that these derived 
efficiencies always exceed the largest efficiency achievable in converting
rest mass into photons around maximally rotating black holes ($43\%$). 
Although this is not a strict theoretical upper limit (models approaching
$100\%$ have been proposed, Lazzati {\it et al}., 2000), still 
values derived from observations are so large to throw doubts on the validity 
of the present version of internal shock model. In fact, 
the maximum efficiency that can be reached with this model is of the
order of 20\% (Guetta, Spada and Waxman 2001a) but only under {\it ad hoc}
assumptions: for random distributions in the wind properties,
the efficiency cannot exceed $\approx .005$ (Spada, Panaitescu, Meszaros 
2000). Perfectly efficient models without internal shocks can be
conceived (Lazzati, Ghisellini, Celotti and Rees 2000), but they
cannot account for bursts' spectra (Ghisellini, Lazzati, Celotti, Rees 2000). 

For this reason,
Kumar and Piran (2000) have proposed an ingenious modification of the
fireball spectra, whereby {\it hot spots} in the emitting surface occasionally
dominate the emission from the whole GRB. These hot spots are not
typical of the whole surface; when our line of sight chances through one
hot spot, we claim to see a burst, but when otherwise, we wouldn't recognize
the event as such because too dim. In this way, they proposed that there be 
a significant number of objects (dissimulated GRBs) escaping our detection, and 
redressing the statistics. The amusing consequence of this is that, while of 
course this implies we would have to {\it decrease} the value of Eq. 
\ref{rategrbs} (thus apparently going in Stecker's direction) there would be 
a substantially higher number of afterglows acting as acceleration sites. 
From Table II of Panaitescu and Kumar, we see that the apparently isotropic
kinetic energy of bursts' {\bf afterglows} is 
\begin{equation}
E_{ag} = 2.5\times 10^{53}\; erg\;.
\end{equation}
This of course is strictly relevant to our acceleration mechanism, because 
as stated above, we are proposing that UHECRs are accelerated at external (= 
afterglow) shocks, so that this is the relevant source of energy to tap. 
Comparison with Eq. \ref{photons} shows indeed that the radiative 
efficiency, defined as $E_{av}/(E_{av}+E_{ag}) \approx 0.6$, very large
indeed. Combining this with Schmidt's rate estimate, we see that there is
a kinetic energy injection rate of
\begin{equation}
\label{afterglow}
\dot{\epsilon}_{ag} = 1.3\times 10^{44}\; erg\; Mpc^{-3}\; yr^{-1}\;,
\end{equation}
comparable with Eq. \ref{rategrbs}. If the argument by Kumar and Piran 
(2000) is right, Eq. \ref{afterglow}
is to be increased. In fact, the kinetic energy per burst in the afterglow is 
derived from very late observations, when the hot spots are not working
any longer (Kumar and Piran 2000), but the rate is to be increased to account
for all bursts for which our line of sight does not meet a hot spot. This
factor is estimated by Kumar and Piran as roughly a factor of $100$, 
bringing Eq. \ref{afterglow} in agreement with Eq. \ref{uhecr}, with a
vengeance. This incompleteness factor can be estimated also as follows:
conventional models (Spada, Panaitescu and Meszaros 2000) find a radiative
efficiency of $\approx 0.005$; observations give $\approx 0.6$, so the hot 
spots bias our statistics by a factor of $\approx 100$. 

\section{Doublets and triplet} 
 
One argument often used against GRBs as sources of UHECRs (and, in general, 
against all explosive sources) is the presence of doublets and triplet 
in the data from AGASA (Takeda {\it et al.}, 1999, 
Nagano and Watson 2000, Hayashida {\it et al.}, 1997),  
where, occasionally, the lower energy  
event precedes the higher energy one. The argument goes that, since the 
time delay with respect to photons is a monotonically decreasing function 
of particle energy, no impulsive source can have produced these clustered 
events, while obvioulsy no such problem exists with continuous sources.  
In the light of the present disagreement between the observational results 
of HiRes and AGASA, one may wonder whether the energy estimates 
even of the events in the clusters are correct. We shall not take this view, 
but accept the existence of the doublets and triplet as established, 
despite the {\it caveat} above, and show that the purported inconsistency 
of the explosive models is not water tight.The argument in fact 
is an oversimplification, and neglects the fact that the time 
delay is a statistically distributed quantity, with both a mean and a  
variance, which we now compute to allow a further discussion.  
 
We take as a model for particle motion a Langevin--like equation 
\begin{equation} 
\label{motion} 
\frac{d\vec{p}}{dt} = m \gamma \frac{d\vec{v}}{dt}= e \frac{\vec{v}}{c} 
\wedge \vec{B} 
\end{equation} 
where, that is, we assume $\vec{B}$ to be a Gaussian field with zero  
mean and known self--correlation function. Initially, we behave
as if $\vec{B}$ were known exactly; we
also use the assumption that the particle Larmor radius $R_L$ is large  
with respect to the typical distances over which the field is correlated,  
say $r_c$: $r_c/R_L \ll 1$, and return later to discuss this approximation.  
We take $z$ as the axis of the initial motion, and decompose in approximate 
form the equations of motion as 
\begin{equation} 
m\gamma\frac{dv_z}{dt}\approx e \frac{v_x}{c} B_y - e B_x \frac{v_y}{c} \;;\; 
m\gamma\frac{d\vec{v}_\perp}{dt} \approx e B_x \hat{y}-eB_y\hat{x}\;. 
\end{equation} 
Here we have assumed that the particle's deflections from the unperturbed 
trajectory are small (which means that $v_z \approx c, v_x\approx v_y\ll c$) 
and thus the equations are valid to smallest 
significant order only; obviously, the fields are to be computed along the unperturbed 
trajectory, $z$. For a source located at distance $D$ from the observer, 
we find 
\begin{equation} 
\bigtriangleup y = \frac{e}{m\gamma c}\int_0^D d\!Z \int_0^Z B_x(z) d\!z 
\;;\; 
\bigtriangleup x =-\frac{e}{m\gamma c}\int_0^D d\!Z \int_0^Z B_y(z) d\!z 
\end{equation} 
and inserting this into the equation for the $z$ component we find 
\begin{eqnarray} 
\label{timedelay} 
\bigtriangleup z= -\frac{e^2}{m^2\gamma^2 c^4}\int_0^D d\!Z \int_0^Z d\!z 
\int_0^z d\!y \left(B_x(z) B_x(z+y)+B_y(z)B_y(z+y) \right) = \nonumber \\ 
-\frac{e^2}{m^2\gamma^2 c^4} \int_0^D (D-z) d\!z 
\int_0^z d\!y \left(B_x(z) B_x(z+y)+B_y(z)B_y(z+y) \right) 
\end{eqnarray} 
where the last identity has been obtained by means of an integration by parts.  
We see that $\bigtriangleup y, \bigtriangleup  
x$, are linear in the fields, and thus are distributed with zero mean, while the 
time delay $\bigtriangleup z$ (which is, by definition, the departure from  
an unperturbed motion with $\gamma \gg 1$, and thus equals the time delay  
with respect to photons), is quadratic in the fields and in the particle charge. 
Because of this, the time delay has a non--zero 
mean, which we now compute by means of the above equation: we obtain 
\begin{equation} 
\label{timed}
<\bigtriangleup z> = -\frac{e^2}{m^2 \gamma^2 c^4}\int_0^D (D-z) d\!z
\int_0^z 2 B_{xx}(y) d\!y \;. 
\end{equation} 
Here $B_{xx}(s) = <B_x(\vec{x}) B_x(\vec{x}+s\hat{z})>$ is the rms 
(since we assumed it to have zero mean) of one component of the field 
pependicular to the line of separation, averaged over all field
configurations; we also used isotropy to set $B_{xx} = B_{yy}$, and 
homogeneity to drop the dependence of $B_{xx}$ on $\vec{x}$. 
We show in the Appendix that, under very reasonable physical assumptions, 
\begin{equation}
B_{xx}(s) 
= \frac{1}{3} B_0^2 \; g(\frac{s}{r_c},\alpha)
\end{equation}
where 
\begin{equation}
g(x,\alpha) \equiv \frac{3\int_0^\infty y^{-\alpha} f(y) d\!y 
(\frac{\sin yx}{y x} + \frac{\cos yx}{(y x)^2} - \frac{\sin yx}{(yx)^3})}
{2\int_0^\infty y^{-\alpha} f(y) d\!y }
\end{equation}
For $\alpha = 5/3$ we would have the well--known Kolmogorov--Okubo law. 

When we insert the above into Eq. \ref{timed}, we see that we have to 
compute the integral
\begin{equation}
\int_0^z d\!y g(y/r_c,\alpha)\;.
\end{equation}
In the above, $z$ is the ($\approx$ cosmological) distance travelled 
by the particle; thus we expect $z \gg r_c$, since, as remarked in the 
Appendix, $z$ is of the order of the radius of the Univers today, while
$r_c$ is of the order of the radius of the Universe at the time of the
generation of the magnetic field. We can then approximate
\begin{equation}
\label{appr}
\int_0^z d\!y g(y/r_c,\alpha) \approx \int_0^\infty d\!y g(y/r_c, \alpha) 
= \frac{\pi r_c}{4}
\frac{3\int_0^\infty y^{-\alpha-1} f(y) d\!y}
{2\int_0^\infty y^{-\alpha} f(y) d\!y } = 
\frac{3\pi}{8} (\alpha-1) r_c
\equiv r_c q(\alpha)\;.
\end{equation}
where we used Eq. \ref{weight}. $q(\alpha)$ is illustrated in Fig. 3. 
Plugging the above into Eq. \ref{timed} we find
\begin{equation}
\label{timedelayfinal}
<\bigtriangleup z> \approx -\frac{q(\alpha) e^2 B_0^2 D^2 r_c }{3 m^2 \gamma^2 c^4}
\;.
\end{equation}
{}

We are interested not just in the mean values, but also in the rms deviations from 
the mean. For the directions perpendicular to the $z$ axis, we easily find with a 
computation analogous to the above 
\begin{equation} 
\sqrt{<(\bigtriangleup y)^2> + <(\bigtriangleup x)^2>} =  
\sqrt{\frac{8 q(\alpha)}{3}} 
\frac{ e B_0 r_c D}{ m\gamma c^2}\left(\frac{D}{r_c} \right)^{1/2}\;. 
\end{equation} 
Two particles arriving on the same experimental apparatus will, in average, be 
separated by $\sqrt{2}$ of the above, and this can be expressed in units of $r_c$: 
\begin{equation} 
\label{separation} 
\delta n= \sqrt{\frac{q(\alpha)}{3}} \frac{4 e B_0 D}{ m \gamma 
c^2}\left(\frac{D}{r_c} \right)^{1/2}\;. 
\end{equation} 
This adimensional quantity $\delta n$ is the number of average correlation  
lengths which 
separate two average particles reaching the same experiment. It essentially 
measures whether two particles of the same energy, 
have travelled through the same patches of 
magnetic field: this is so, whenever $\delta n \ll 1$, or otherwise when  
$\delta n \gtrsim 1$.  
 
In the case $\delta n \ll 1$, we may reasonably assume that the  
particles of the same energy emitted simulataneously in an explosive event, 
will arrive at the experiment also simultaneously, with the time--delay  
derived above. In the case $\delta n \approx 1$ or greater, since these  
particles will have travelled 
through patches of uncorrelated magnetic field, we cannot assume any longer 
that the particles will arrive simultaneously; a more reasonable assumption 
is that their time delays be distributed according to the statistics implicit 
in Eq. \ref{timedelay}. We will present the full statistics elsewhere, and will 
limit ourselves here to derive the rms deviation of the time delay around 
its mean. We proceed as follows: using Eq. \ref{timedelay} we find 
\begin{eqnarray} 
(\bigtriangleup z)^2 = \frac{e^4}{m^4 \gamma^4 c^8} \int_0^D d\!z \int_0^D d\!z' 
(D-z) (D-z') \int_0^z d\!s \int_0^{z'} d\!s' \nonumber \\ 
(B_y(z)B_y(z+s)+B_x(z)B_x(z+s)) 
(B_y(z')B_y(z'+s')+B_x(z')B_x(z'+s'))\;. 
\end{eqnarray} 
We now average the above over all field configurations; to do so, we use 
Wick's theorem since we assumed $\vec{B}$ to be a Gaussian field: 
\begin{eqnarray} 
<(B_y(z)B_y(z+s)+B_x(z)B_x(z+s))(B_y(z')B_y(z'+s')+B_x(z')B_x(z'+s'))> = \nonumber \\ 
2(2 B_{xx}(s)B_{xx}(s') + B_{xx}(|z'-z|)B_{xx}(|s-s'|) +  
B_{xx}(|z-z'-s'|)B_{xx}(|z+s-z'|)\;. 
\end{eqnarray} 
We recognize in the first term to the right of the equality the term which leads 
to $(<\bigtriangleup z>)^2$; since it is our aim to derive the rms in the time delay, 
we drop it from now on. The second term on the rhs of the above can be seen to vanish 
(or more precisely, to give exponentially small terms when $B_{tt} \propto  
\exp(-D/r_c)$), and will thus be neglected. Putting together all of the 
above, we find 
\begin{equation} 
\Delta_{rms} \equiv <(\bigtriangleup z)^2>-(<\bigtriangleup z>)^2 =  
\frac{ q^2(\alpha) e^4 r_c^2 B_0^4 D^4}{18 m^4 \gamma^4 c^8} 
\end{equation} 
and we now define a convenient quantity as 
\begin{equation} 
\label{delta} 
\Delta \equiv \frac{\sqrt{\Delta_{rms}}}{<\bigtriangleup z>} = \frac{1}{\sqrt{2}}\;, 
\end{equation} 
which is the result we were searching for. It is worthwhile remarking that 
this is independent of all quantities, so long as our approximations (Eq. \ref{appr})
are satisfied. 
 
We now may begin our discussion by remarking that propagation in the IGM clearly  
satisifies the hypothesis under which these results were derived, {\it i.e.},  
$r_c/R_L \ll 1$ and homogeneity, isotropy, and gaussianity (HIG, from now on) 
of the field (see Grasso and Rubinstein 2001 for a review),  
so that the computation is immediately applicable. We find for  
Eq. \ref{separation}: 
\begin{equation} 
\delta n = 1 \frac{B_0}{2\times 10^{-11}\; G}\frac{D}{100\; Mpc} 
\frac{3\times 10^{10}}{\gamma}\left(\frac{D}{100 r_c}\right)^{1/2} 
\end{equation} 
showing that, for very reasonable values of the parameters involved, particles 
which reach Earth experiments cross regions with uncorrelated magnetic field. 
In this case, these  
particles' arrival times are spread out around their mean value (Eq.  
\ref{timedelay}) with an rms of order $\Delta$; if doublets and triplets  
include particles with energies different by a factor $p$, their mean  
arrival times differ by a factor $1-1/p^2$, and this corresponds to a splitting 
of $(1-1/p^2)/\Delta$ times the rms arrival time, for the lower energy ones. 
For typical observed values of $p \approx 2$, we find that the separation of 
the means in arrival times is only $3\sqrt{2}/4\approx 1$, implying a  
separation of $1\sigma$ only. This shows  
that there must be a superposition between the higher energy ones (assumed here 
for sake of argument to be travelling without dispersion in arrival times) 
and the lower energy ones (assumed instead to be spread out). Also, it is 
worthwhile to remark that the flux suppression at the lower energy due to  
statistics ($\approx e^{-2}$, as will be shown in aforthcoming paper) is nearly  
exactly compensated for by larger number 
of low energy particles: for a spectrum going like $\approx E^{-3}$ at Earth, 
the product of the two, $e^{-2}p^3 \approx 2$, again in agreement with observations. 
 
However, besides crossing the IGM, any particle reaching the Earth must also cross 
the Local Supercluster, and its associated magnetic field. Since the Local  
SuperCluster is obviously not a virialized structure, we may reasonably  
assume that HIG holds again. The current value of  
$B$ is rather uncertain. If we take $D = 10\; Mpc$, $ B \approx3\times 10^{-8}\; G$ and 
$r_c = 1 \; Mpc$ as discussed by Blasi and Olinto (1999), we find $r_c/R_L = 
1/3$, so that the above computations are still (marginally) applicable; also 
$\delta n \approx 10$. A more recent determination (Vallee 2002) has $B \approx 
1\;\mu G$, and $r_c = 100\; kpc$, so that again $r_c/R_L = 1$, and the  
computation is marginally applicable; for these parameter values, $\delta n \approx 
100$.  
 
The bottom line of this argument is that, either in the IGM (for reasonable parameter 
values) or in the Local Supercluster (for currently favored, but still very  
uncertain parameter values), the protons which reach the same experimental apparatus 
on Earth will have travelled through disconnected patches of magnetic field, and 
their time of arrivals will be spread around the mean by their large (Eq.  
\ref{delta}) variance, and will mostly wash out the strict time ordering which  
is to be expected in the case of particles of (formally) infinite energy. 

A different, and important question, is whether the statistics implicit in Eq. 
\ref{timedelay} is consistent with the several instances of doublets, and triplet,
observed, and the number of observed inversion in the ordering of arrivals. 
A full answer to 
this question requires working out the full statistics implicit in Eq. 
\ref{timedelay}, and the comparison with the real data; the details of this
work lie outside the scope of this work, and will be presented elsewhere. 
But it is nonetheless worth remarking that, on the basis of the arguments
presented above, we expect roughly equal occurrences of multiplets with
lower (higher) energy particles preceding higher (lower) energy particles. 
From Table 2 of Takeda {\it et al}. (1999) we see that this is precisely the case:
in multiplets C1,C4,C5 the higher energy particle precedes the lower energy
one, and the reverse occurs in multiplets C2 and C3. 

\section{Comparison with other work} 
 
Simultaneously with, and independently of Vietri (1995), two other papers  
(Milgrom and Usov 1995, Waxman 1995) proposed that UHECRs were accelerated 
in GRBs, but, while Milgrom and Usov did not propose a specific acceleration 
mechanism, Waxman (1995) proposed that Type II Fermi acceleration at the 
subrelativistic internal shocks could account for the acceleration of UHECRs.

It is however still controversial whether the internal shocks can 
account for the GRB emission in the prompt phase. Spectra of GRBs display  
both a universal break feature at $\approx  
200 \; keV$ (Preece {\it et al.}, 2000) and a wildly variable low energy  
spectral index (Crider {\it et al.}, 1997, Preece {\it et al.}, 1998,  
Ghirlanda, Celotti and Ghisellini 2002), which have proved difficult to  
reproduce theoretically: to wit, look at Fig. 1 of Meszaros and Rees (2000).  
In particular, it seems that pure synchrotron emission is incapable of 
explaining at least some (but not all! Tavani 1996, 1997)
of the bursts' spectral properties  
(Ghisellini, Celotti, Lazzati 2000). Possible alternatives have been  
discussed (Pilla and Loeb 1998, Ghisellini and Celotti, 1999, Panaitescu  
and Meszaros 2000) but none of them is widely credited with being able to 
explain the above--mentioned properties of the observed bursts.  
These difficulties should be contrasted with the success of the  
external shock model, which is widely credited with the explanation 
of the properties of the afterglow (Piran 2000).

Since internal shocks are Newtonian (Waxman 2002), we do not expect  
the acceleration time scales to be short enough: the large average energy  
gain discussed above ($G\approx 2$, Gallant and Achterberg 1999) only applies  
to relativistic shocks, and the ability of internal shocks to provide  
UHECRs in the short times allotted remains to be proved. This is  
especially so, since Waxman (1995, 2002) is appealing to the much less 
efficient Type II acceleration, rather than the standard Type I. But,
even if TypeII acceleration at a Newtonian shock should prove rather fast,
still numerical 
simulations (see {\it e.g.} Meli and Quenby 2002) show explicitly that  
acceleration at relativistic shocks is much faster than Type I acceleration 
at Newtonian shocks, and this, in its own turn, is much faster than Type II
acceleration (again at a Newtonian shock). Thus we may always expect 
Type I acceleration at external shocks to dominate Waxman's assumed mechanism.

There is no obvious reason to suspect that particle acceleration at 
internal shocks should be particularly effective: as emphasized by Waxman (2002), 
internal shocks are Newtonian, and thus a strong counterpressure 
due to particles will result in a number of instabilities (Ryu, Kang and Jones 1993, 
and references therein) 
and in the smoothing out of the shock. The major consequence of this would 
be the suppression of the lower energy electrons which are supposed to  
radiate away the winds' internal energy, and thus the muting of the burst. 
Acceleration of UHECRs at external shocks, instead, suffers no such problem 
\footnote{We are indebted to P. Blasi for pointing this out to us.}: since 
it takes only a deflection by an angle ${\cal O}(1/\Gamma)$ to allow the  
shock to overtake a particle in the upstream region, as $\Gamma\rightarrow 
\infty$ the linear momentum communicated to the upstream section vanishes 
as $1/\Gamma^2$, and no counterpressure/smoothing of the shock/instabilities ensue.

The simple energy budget 
favors acceleration at external shocks. The apparently large radiative  
efficiency of bursts, contrary to all models for internal shocks 
(Spada, Panaitescu, Meszaros 2000), has prompted Kumar and Piran (2000) 
to propose an ingenious model to salvage internal shocks. According to 
them, hot spots in the shock surface  
strongly bias our detection abilities in their favor, and make it look 
like the overall radiative efficiency approaches unity; the many 
bursts without hot spots aiming at us, which are necessary to redress 
the statistic, would not be detected. In this model, the {\bf true} radiative  
efficiency in the internal shock phase would be about $0.01$ of what is 
observed, and there should be about $1/0.01 = 100$ more GRBs than we detect. 
But what this means is that there would be $100$ times as many external  
shocks from which to draw UHECRs, because external 
shocks, for which beaming effects are smaller, (and thus also the hot spot 
mechanism would not be operational, Kumar and Piran 2000) suffer   
through no such criticism (Vietri 1998a). In other words: either one believes 
the largish ($\approx 0.18$ for GRB 970508) radiative efficiencies, in which  
case it is hard to believe in the internal shock model; or, one tries to salvage 
the internal shock model, in which case the energy balance is dominated  
by a large margin by the kinetic energy of the afterglow phase, which can be  
tapped by acceleration of UHECRs at external shocks.

Apart from these objections, it is not easy to conceive at this 
point an observational test capable of distinguishing between the two models. 
The only hope appears to be the production of high energy neutrinos which  
must accompany the {\it in situ} acceleration 
of particles; occasionally, in fact, UHECRs will produce pions (and  
then neutrinos) through collisions with photons, in the moderately  
photon rich environment provided by the post--shock shells. 
If UHECRs are accelerated at internal shocks, the neutrinos  
thusly produced will arrive at Earth simultaneously with the photons 
of the burst proper and will have an energy $\sim 10^{15} eV$  
(Waxman and Bachall 1997, Guetta Spada Waxman 2001b).  
If UHECRs are accelerated at external  
shocks they will arrive at Earth simultaneously with the photons 
of the afterglow and will have a higher energy, $\gtrsim 10^{17}$ 
(Vietri 1998a, 1998b).  

\section{Conclusions}
 
In short, an analysis of the objections raised against acceleration 
of UHECRs at the external shocks of GRBs shows that none of them is, at the
time of writing, dangerous for the model.

The strongest objection levied so far is that  
due to Gallant and Achterberg (1999) that the highest energies of particles  
accelerated at external shocks is smaller than surmised in Vietri (1995); 
while in fact we pointed out that this objection is model--dependent, so 
is the reply, and ultimately the success of the scenario depends upon 
whether models in which a GRB explodes in a PWB (as discussed by K\"onigl  
and Granot 2002) or some magnetic--field--rich environment (Blandford and 
Begelman 1998, Usov 1992, Drenkhahn and Spruit 2002, Duncan and Thompson
1992, Dai and Lu 1998, Blackman and Yi 1998) are correct, or not. 

The strongest success of the model is once again the close 
agreement between the energy release in $\gamma$--ray photons and that
necessary to account for UHECRs at Earth: to wit, compare Eqs. 
\ref{uhecr} and \ref{afterglow} (including the whole discussion,
in Section 4). This agreement requires slightly different production 
efficiencies for photons and high energy particles, but is otherwise 
unique to this model. 

It is also worth remarking that 
much information is contained in the observed properties of UHECRs: as 
noted above, in fact, multiplets contain information on the properties
of the magnetic field along the line of sight which is peculiar to
strongly time--varying sources, and which, coupled with other observed
features such as the spectrum and the overall number of sources on the 
plane of the sky, can put the model through a severe test. A detailed
analysis of this kind will be presented elsewhere.

It is a pleasure to acknowledge extremely helpful conversations with P. Blasi. 
{} 

\section*{Appendix}

We wish to determine here the correlator for the Gaussian field:
\begin{equation}
B_{ij}(\vec{s}) \equiv <B_i(\vec{x})B_j(\vec{x}+\vec{s})>;.
\end{equation}
We start out by considering the Fourier transform of $B_{ij}$:
\begin{equation}
\label{fourier}
b_{ij}(\vec{k}) \equiv \int B_{ij}(\vec{s}) e^{-\imath \vec{k}\cdot\vec{s}}
d^3\!s\;.
\end{equation}
Because of homogeneity, isotropy, and Maxwell's equation $\bigtriangledown\cdot
\vec{B} = 0$ one finds (Landau and Lifshitz 1987)
\begin{equation}
\label{symmetry}
b_{ij}(\vec{k}) = \frac{2\pi^2 E(k)}{k^2}(\delta_{ij}-\frac{k_i k_j}{k^2})
\end{equation}
where $E(k)$ is the mode energy distribution. If $E(k)$ were to follow
a pure Kolmogorov--Okubo law, we would have $E(k) = A k^{-5/3}$, but 
we have to modify this in two ways. First, the magnetic field outside of 
galaxy clusters cannot be due to the usual small scale turbulence that 
we are accustomed to in the laboratory, and thus the exponent in the above
law need not equal $5/3$. Second, if phase transitions at an early epoch
(Grasso and Rubinstein 2001) are responsible for the generation of the
magnetic field outside galaxy clusters, the field cannot be correlated 
over lengthscales exceeding the horizon scale at field generation. So
we need to introduce a cutoff $f(r_c k)$ for all modes with $k < 1/r_c$, 
for some $r_c$ which we shall assume known from now on. We thus take
\begin{equation}
\label{smoothed}
E(k) = A k^{-\alpha} f(r_c k)\;.
\end{equation}
Strictly speaking, we do not know $f(x)$ either; for computational purposes,
in the following we shall take 
\begin{equation}
\label{weight}
f(x) \equiv e^{-1/x}\;.
\end{equation}
Taking $\vec{s} = s \hat{z}$, we now compute $B_{xx}(s) = B_{yy}(s)$. From the
inverse of Eq. \ref{fourier} we find
\begin{equation}
B_{xx}(s) = \int b_{xx} e^{\imath\vec{k}\cdot\vec{s}}\frac{d^3\!k}{(2\pi)^3}
\end{equation}
which, using Eqs. \ref{smoothed} and \ref{symmetry} yields
\begin{eqnarray}
B_{xx}(s) = \frac{\pi A}{4}\int k^{-\alpha} f(r_c k)(\cos^2\theta+\sin^2\theta
\sin^2\phi) e^{\imath k s \cos\theta}d\!k\;d\!\phi\; d\!\cos\theta = \\ \nonumber
\frac{\pi^2 A}{2} \int_0^\infty k^{-\alpha} f(r_c k) d\!k \int_{-1}^1
\frac{1+\cos^2\theta}{2} e^{\imath k s \cos\theta} d\!\cos\theta\;.
\end{eqnarray}
We find
\begin{equation}
B_{xx}(s) = \pi^2 A \int_0^\infty k^{-\alpha} f(r_c k) d\!k 
\left(\frac{\sin ks}{ks}+\frac{\cos ks}{(ks)^2}-\frac{\sin ks}{(ks)^3}\right)\;.
\end{equation}
When $s\rightarrow 0$, because of isotropy obviously $B_{xx}\rightarrow 
B_0^2/3$, where $B_0$ is the rms field. Letting $s\rightarrow 0$ in the above 
equation, we find 
\begin{equation}
B_0^2 = 2 \pi^2 A \int_0^\infty k^{-\alpha} f(r_c k) d\!k
\end{equation}
which allows us to rewrite 
\begin{equation}
\label{final}
B_{xx}(s) = \frac{1}{3} B_0^2 \; g(\frac{s}{r_c},\alpha)\;\;;\;\;
g(x,\alpha) \equiv \frac{3\int_0^\infty y^{-\alpha} f(y) d\!y 
(\frac{\sin yx}{y x} + \frac{\cos yx}{(y x)^2} - \frac{\sin yx}{(yx)^3})}
{2\int_0^\infty y^{-\alpha} f(y) d\!y }
\end{equation}
and similarly for $B_{yy}$.

{} 

\clearpage
 
 \begin{figure} 
 \centering 
 \noindent 
 \includegraphics[width=12cm]{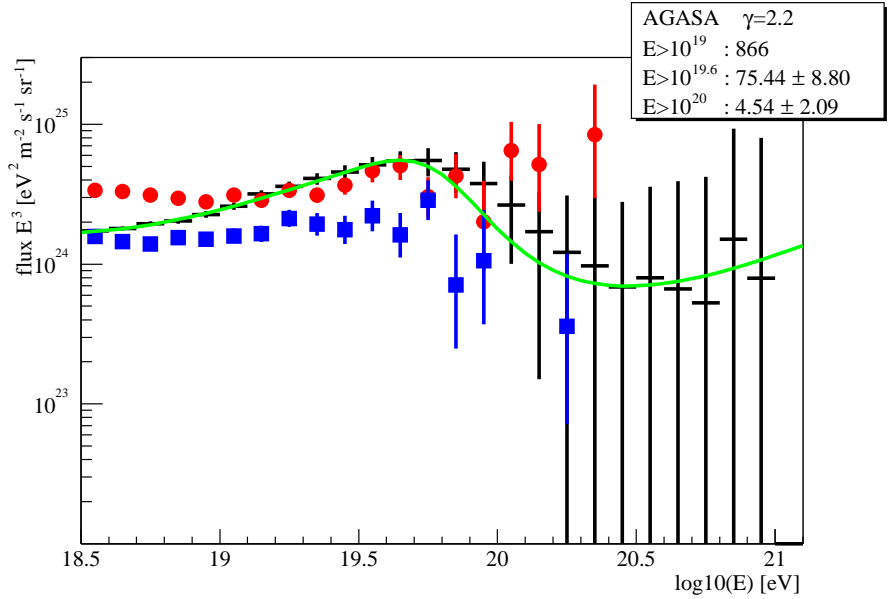} 
 \caption{\label{fig1}  
 Comparison of expected and observed spectra of UHECRs at Earth. Solid
 line, analytical computation of the spectrum. Crosses with bars, 
 numerical simulations of the AGASA data. Circles with bars, AGASA
 data. Suqares with bars, HiRes data. The bursts' redshift distribution is 
 Porciani and Madau's (2001) SFR2. The injection spectrum extends over 
 the range given by Eq. \ref{range}; the injection spectral slope
 is $\gamma = 2.2$.}
 \end{figure}
\newpage
 
 \begin{figure} 
 \centering 
 \noindent 
 \includegraphics[width=12cm]{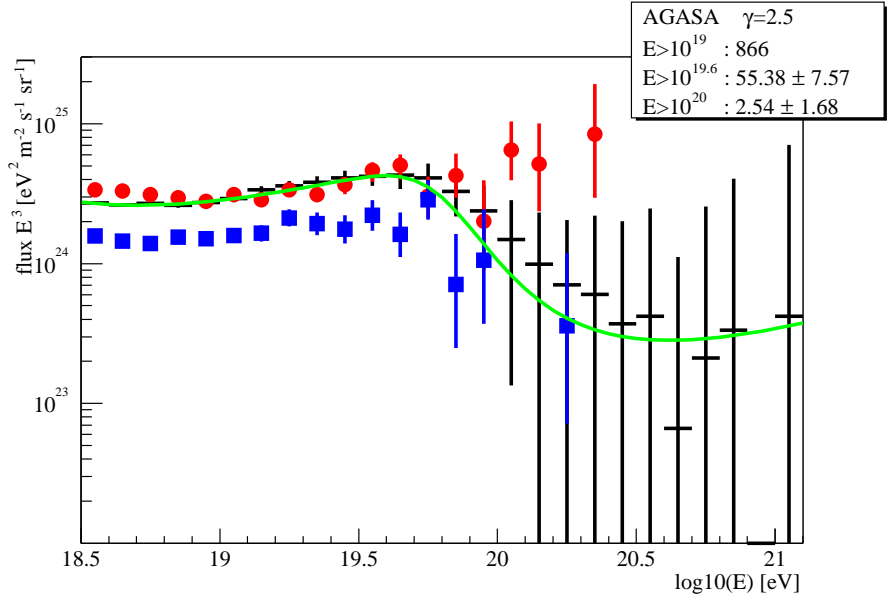} 
 \caption{\label{fig2}  
 Same as Fig.1, except that $\gamma = 2.5$.}
 \end{figure}
 
 \newpage 
  
 \begin{figure} 
 \centering 
 \noindent 
 \includegraphics[width=12cm,angle=-90]{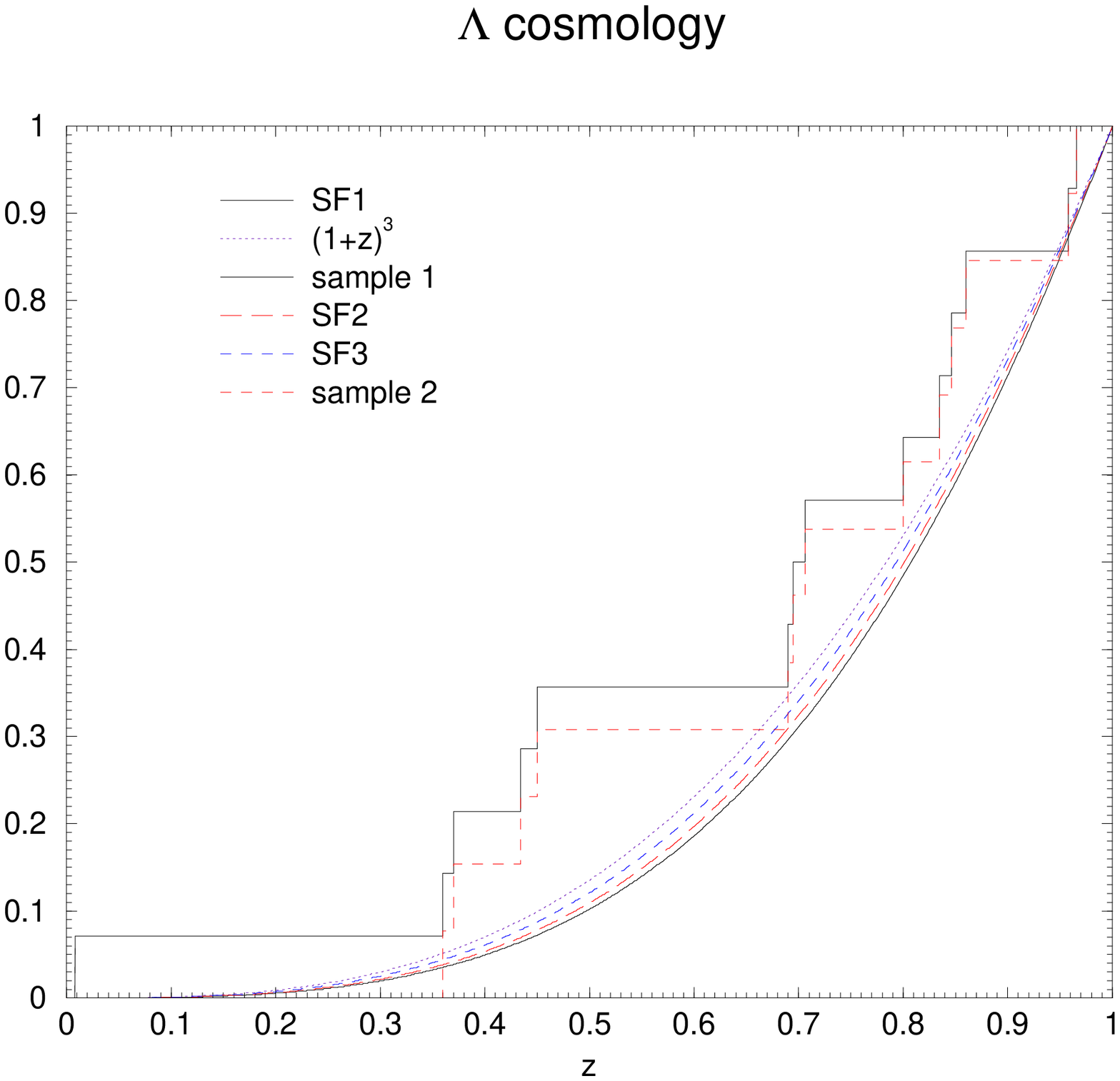} 
 \caption{\label{fig3}  
 The observed cumulative distribution function in redhsift, for all GRBs with 
 $z \leq 1$. The two observed samples exclude (Sample 1) or include (Sample 2) 
 the low redshift burst GRB 980425. 
 These observed distributions are compared with theoretical distributions, 
 obtained for three different star formation histories, taken from Porciani 
and  Madau 2001: SF1 (continuous curve), SF2 (long-dashed curve), SF3 (dashed 
curve). The dotted curve shows a non--evolving distribution $\propto (1+z)^3$.} 
\end{figure} 
 
\end{document}